\documentstyle[graphicx]{europhys}

\def\etal{{\hbox{{\tenit\ et al.\/}\tenrm :\ }}}

\def\And{{\rm and\ }}

\def\stars{\bigskip\centerline{***}\medskip}

\newif\ifboo \boofalse

\def\Review#1{\boofalse{\it #1},}
\def\Name#1{{\sc #1},}
\def\Vol#1{\ifboo Vol. {\bf #1}\else{\bf #1}\fi}
\def\Year#1{\ifboo #1\else(#1)\fi}

\def\Page#1{\ifboo {\rm p. #1}\else{\rm #1}\fi}

\def\etal{ \textit{et al.} : }

\def\eurologo{{\rm EURO-\LaTeX\ (does not imply 
endorsement by \sc{Europhysics Letters})}}

\makeatletter
\def\vereq#1#2{\lower3pt\vbox{\baselineskip1.5pt \lineskip1.5pt
\ialign{$\m@th#1\hfill##\hfil$\crcr#2\crcr\sim\crcr}}}
\def\agt{\mathrel{\mathpalette\vereq>}}
\def\alt{\mathrel{\mathpalette\vereq<}}
\makeatother

\newcommand{\kB}{k_{\mathrm{B}}}
\newcommand{\kT}{\kB T}
\newcommand{\wD}{w_{\mathrm{D}}}
\newcommand{\f}{{\bf f}}
\renewcommand{\v}{{\bf v}}
\newcommand{\r}{{\mathbf r}}
\newcommand{\vpll}{\v^{||}}
\newcommand{\rc}{r_{\mathrm{c}}}
\newcommand{\amp}{A}
\newcommand{\etaK}{\eta_{\mathrm{K}}}
\newcommand{\etaD}{\eta_{\mathrm{D}}}
\newcommand{\lambdaMBE}{\lambda_{\mathrm{MBE}}}
\newcommand{\ncoll}{n_{\mathrm{coll}}}
\newcommand{\vrms}{v_{\mathrm{rms}}}
\newcommand{\gammaC}{\gamma_{\mathrm{c}}}
\newcommand{\rhoC}{\rho_{\mathrm{c}}}
\newcommand{\var}{{\mathrm{var}}\,}
\newcommand{\Sc}{Sc}
\newcommand{\eqref}[1]{(\ref{#1})}

\begin{document}
\euro{}{}{}{}
\Date{}
\shorttitle{A. J. MASTERS \etal KINETIC THEORY FOR DPD ETC.}
\title{Kinetic theory for dissipative particle dynamics:\\
the importance of collisions}
\author{A. J. Masters\inst{1} \And P. B. Warren\inst{2}}
\institute{
\inst{1}Department of Chemistry, University of Manchester,\\
Oxford Road, Manchester, M13 9PL, UK.\\
\inst{2}Unilever Research Port Sunlight Laboratory,\\
Quarry Road East, Bebington, Wirral, L63 3JW, UK.}
\rec{}{}
\pacs{
\Pacs{02}{70Ns}{Molecular dynamics and particle methods}
\Pacs{05}{40$+$j}{Fluctuation phenomena, random processes and Brownian motion}
\Pacs{66}{20$+$d}{Viscosity of liquids, diffusive momentum transport}}
\maketitle

\begin{abstract}
Kinetic theory of dissipative particle dynamics is developed in terms of
a Boltzmann pair collision theory. The kinetic transport coefficients
are computed from explicit collision integrals and compared favourably
with detailed simulations. Previous theory is found to correspond to a
weak scattering limit, or Vlasov theory, and previously reported
discrepancies with simulations are thereby resolved. In the large
dissipation limit, we find qualitatively new scaling properties for the
transport coefficients.
\end{abstract}

Dissipative particle dynamics (DPD) is emerging as an attractive
simulation method in such diverse areas as colloidal suspension
rheology, explicit multiphase flow problems, polymer solution dynamics,
and phase behaviour of block copolymer melts \cite{DPD}. The basic DPD
fluid consists of a large number of identical particles interacting by
pairwise soft repulsive, dissipative and random forces \cite{HK}. The
soft repulsions typically correspond to an interparticle potential, $U =
\amp(1-r/\rc)^2/2$, where $r$ is the distance between a pair of
particles, $\rc$ sets the range and $\amp$ sets the amplitude. The
dissipative forces are similarly specified by $\f = -m\gamma\wD(r)\vpll$
where $m$ is the particle mass, $\gamma$ (with units of inverse time)
sets the overall dissipation rate, the weight function is usually
$\wD(r)=(1-r/\rc)^2$, and $\vpll$ is the relative velocity of the
particles projected onto the line joining their centres. Given the
dissipative forces, the random forces are completely determined by a
fluctuation-dissipation theorem apart from an overall amplitude which
determines the temperature $\kT$; their detailed form need not be further
specified \cite{EW,GW}. All forces vanish for $r>\rc$, and are
arranged to conserve momentum locally. Thus hydrodynamics is recovered
at longer length and time scales just as in a molecular fluid.

For the applications cited above, the basic interactions are typically
augmented by extra bond constraints, different types of particles, and
so on. In nearly all cases though the properties of the basic DPD fluid
are required to calibrate the results against other methods and to
provide an essential bridge back to the real world. Thus there has been
considerable effort in establishing a rigorous theoretical basis for the
method \cite{EW} and in developing kinetic theories for the transport
properties \cite{Esp,MBE}. These studies also open up relatively
unexplored terrain for kinetic theory. For instance the `ideal
dissipative fluid' (basic DPD fluid with $\amp=0$) is probably the
simplest example of a completely structureless fluid with non-trivial
transport properties. Dimensional analysis indicates that the transport
properties of this fluid are completely determined by a dimensionless
dissipation rate, $\gamma\rc(\kT/m)^{1/2}$, and number density,
$\rho\rc^3$, where $\rho$ is the number of particles per unit volume. In
what follows, we shall use units in which $m=\rc=\kT=1$, so that the rms
velocity of particles is $\vrms=\sqrt{3}$ for instance. In these units,
the general DPD fluid is completely specified by $\gamma$, $\rho$ and
$\amp$.

Our starting point is the kinetic theory developed by Marsh, Backx and
Ernst (MBE) \cite{MBE}. They identify the principal time scales in the
ideal dissipative fluid, and provide explicit expressions for the
transport coefficients. For
instance, the (normalised) MBE velocity autocorrelation function (VACF)
is predicted to decay exponentially, $\phi(t)=e^{-t/\tau}$, with a decay
rate
\begin{equation}
\tau^{-1}=\frac{1}{3}\int\!\rho\,d^3\r\,\gamma\wD(r)
=\frac{2\pi\gamma\rho}{45}.\label{taueq}
\end{equation}
For later use we define $\lambdaMBE=\tau\vrms=45\sqrt{3}/2\pi\gamma\rho$ to
be a representative mean free path in the MBE theory. The self diffusion
coefficient in the MBE theory, given our choice of units above, is
simply $D=\tau$. The viscosity in the MBE theory has two contributions:
a kinetic contribution, $\etaK=\rho D/2$, and a dissipative contribution
arising directly from the dissipative forces, $\etaD =
(\gamma\rho^2/2)\, \int\!d^3\r\,r^2\,\wD(r) = 2\pi\gamma\rho^2/1575$.
In simple terms these correspond to the internal friction induced by
particles diffusing across streamlines, and the
drag force that acts between particles on different streamlines.

The MBE theory captures the basic kinetic phenomena in the DPD fluid,
and is highly successful in explaining such apparent paradoxes as a
decrease in viscosity for certain parameter ranges, as the dissipation
rate $\gamma$ is increased. When compared in detail with simulations
though, certain discrepancies were already noted by the original
investigators \cite{MBE}, and were explored more recently by
Pagonabarraga et al \cite{PHF}. Neither set of authors identify the
origin of the discrepancies, although correlation effects are an obvious
candidate. 

To investigate these discrepancies in detail, we carried out a
systematic study of the ideal dissipative fluid by simulation
\cite{sims}. Typical results for the VACF are shown in Fig.~1. We
find that $\phi(t)$ is remarkably insensitive to $\rho$ but depends
significantly on $\gamma$, once a basic scaling prefactor $\rho\gamma$
is removed from the time dependence. For $\gamma\alt5$ the VACF decays
as a single exponential, to an excellent approximation, over two decades
of magnitude (after which the signal becomes swamped by noise). The decay
rate is plotted against $\gamma$ in the inset to Fig.~1(a); it shows a
systematic deviation from the MBE prediction. For $\gamma\agt5$
significant deviations from single exponential decay start to appear,
indicating the early onset of correlation effects. Of course one would
expect correlation effects to be present to some extent for all
$\gamma$, since they lead for instance to the celebrated long time tail
$\phi(t)\sim t^{-3/2}$ \cite{LTT}.

We also studied the two contributions to the viscosity.
Typical results are shown in Fig.~2. Although $\etaD$ appears to be well
captured by the theory, $\etaK$ is again systematically wrong, by as
much as a factor of three (Fig.~2(b)). It too is remarkably insensitive
to density (Fig.~2(a)).

When the VACF is well approximated by a single exponential, significant
discrepancies with theory cannot be ascribed to the correlation effects
hypothesised earlier. Furthermore, the lack of dependence of the kinetic
properties on density is startling. For a long time these results
puzzled us, until we recalled the basic properties of the classic
Boltzmann pair collision theory \cite{Res}. Let us first introduce an
estimate of the number of collisions a particle undergoes before it
loses its memory of its initial velocity. To be specific, define
$\ncoll=\rho\lambda$ where $\lambda$ is the mean free path (note that
the collision cross section $\sim\rc^2=1$). In the MBE theory, this
gives $\ncoll = \rho\lambdaMBE=45\sqrt{3}/2\pi\gamma$; we will use this
estimate to analyse the data. The importance of a parameter proportional
to $\rho\lambdaMBE$ was recently noted by Evans in a different approach
\cite{Evans}. 

Our central argument is that $\ncoll$ and pair collisions in general
are the key to understanding the results. Firstly note that in a pair
collision theory, the transport coefficients $\rho D$ and $\etaK$ have
no dependence on density. Secondly, when the VACF decay rate (Fig.~1)
and $\etaK$ normalised by the MBE prediction (Fig.~2(b)) are examined as
functions of $\ncoll\sim1/\gamma$, a data collapse is found. Thirdly,
both the VACF decay rate and $\etaK$ systematically asymptote towards
the MBE theory in the limit $\ncoll\to\infty$. These observations imply
that the kinetic properties of the fluid are principally determined by
pair collisions ($\ncoll$ in effect), and that MBE theory is actually
only weak deflection theory, valid in the limit where many collisions
are required to decorrelate a particle's velocity.

If pair collisions are the determining factor, the transport properties
ought to correspond to certain collision integrals in a Boltzmann
theory. It is straightforward to show that the standard expressions hold
for the DPD fluid with the minor modification that, for a given impact
parameter and pre-collisional set of velocities, the random force leads
to a distribution of outgoing velocities rather than a single, unique
set as is the case for normal fluids. The collision integrals take the
generic form \cite{Res}
\begin{equation}
\Omega_{D,\eta}=\int_0^1\!2\pi b\,db\,
\int\!d^3\v_{12}\,\phi_0(\v_{12})\,|\v_{12}|\,
\langle F_{D,\eta}(\v_{12},\v'_{12})\rangle
\end{equation}
where $b$ is the impact parameter in units of $\rc$, $\v_{12}$ is the
pre-collisional relative velocity, $\v'_{12}$ is the post-collisional
relative velocity and $\phi_0$ is the Maxwellian distribution function
for relative velocities. The functions are $F_D(\v,\v') = v_x
(v'_x-v_x)$ for the self diffusion coefficient, and $F_\eta(\v,\v') =
v_xv_y (v'_xv'_y-v_xv_y)$ for the kinetic contribution to viscosity. The
angle brackets denote an average over the distribution of out-going
velocities for a given $b$ and $\v_{12}$.

To evaluate these collision integrals, we used a numerical, Monte Carlo
approach. A value of $b^2$ was chosen from a uniform distribution in the
range $[0,1]$ and a value of $\v_{12}$ was taken from the Maxwellian
distribution. We then followed the particle trajectory using the
algorithm of Pagonabarraga et al \cite{PHF}. We noted the outgoing
relative velocity and thus formed the integrand for this particular
trajectory. By averaging over many trajectories we obtained the overall
integral. Typically we used a time step of 0.005 to compute the
trajectory and $10^8$ trajectories to get an precision of better than
$1$\%.

Once the collision integrals were determined, we calculated the
diffusion constant and the viscosity using the standard results (in our
chosen set of units) $\rho D=-4/\Omega_D$ and
$\etaK=-8/\Omega_\eta$. These results are at a first Sonine
polynomial level of approximation. For the hard sphere fluid, for
example, this leads to an error of $<2$\%, compared to the true
Boltzmann value \cite{Res}. To check whether the first Sonine polynomial
approximation was as good for the DPD fluid, we also calculated the
second Sonine polynomial correction. Just as in the hard sphere fluid,
the correction was of the order of 1\% so we believe the results quoted
are excellent approximations to the true Boltzmann values.

The predictions of the Boltzmann pair collision theory turn out to be in
excellent agreement with the simulation results, thus confirming our
premise that collisions are the key. Results for the decay rate of the
VACF (defined by $\tau=D$) are shown as open circles in the inset to
Fig.~1(b). Similar results for $\etaK$ in shown in Fig.~2(b). Moreover,
we have been able to prove that the MBE theory is obtained in the Vlasov
limit \cite{Res}, where each collision only weakly perturbs the
velocities of the colliding particles, thus verifying that the MBE
theory is indeed a weak deflection theory.

Now consider the significance of these ideas for the strong deflection
limit. Since $\ncoll\sim1/\gamma$ in the MBE theory, it appears that for
$\gamma$ sufficiently large, $\ncoll<1$. This is nonsense though,
because at least one collision is needed to decorrelate a particle's
velocity no matter how large $\gamma$ is. Thus for $\gamma$ sufficiently
large there must be a qualitative change in the behaviour. The crossover
occurs at $\ncoll\approx1$ or $\gamma\approx\gammaC =
45\sqrt{3}/2\pi\approx12.4$. Turning the argument around, if
$\ncoll\agt1$ is enforced for all $\gamma$ there must be a change in the
scaling of the mean free path from the MBE result,
$\lambdaMBE\approx\gammaC/(\gamma\rho)$ for $\gamma\alt\gammaC$, to a new
behaviour, $\lambda\approx1/\rho$ for $\gamma\agt\gammaC$. 

This argument implies that $\tau$, $D$ and $\etaK$ should reach a
plateau for $\gamma\agt\gammaC$. This is in marked contrast to the
predictions of both the MBE theory \cite{MBE} and Evans' theory
\cite{Evans}. That a plateau is indeed found is demonstrated for $\etaK$
in the inset to Fig.~2(b). We find a good fit is
$\etaK=0.37(1)+3.60(5)/\gamma$ (the numbers in brackets are estimates of
the error in the final digit). Similarly for the self diffusion
coefficient (data not shown here) we find $D=0.21(1)+2.52(2)/\gamma$. At
present we do not have an explanation for the straight line fits, but we
can use them pragmatically to define crossovers at
$\gamma\approx3.60(5)/0.37(1) = 9.7(3)$ for $\etaK$ and
$\gamma\approx2.52(2)/0.21(1) = 12.0(6)$ for $D$. 
These are satisfyingly close
to $\gammaC\approx12.4$. This conclusion has implications for Schmidt
number, an important dimensionless material parameter defined by
$\Sc=\eta/\rho D$. In the large $\gamma$ (fixed $\rho$) limit, the
viscosity is dominated by $\etaD\sim\rho^2\gamma$ but the above argument
changes the scaling of $D$ from $(\rho\gamma)^{-1}$ to $\rho^{-1}$. This
suggests that $\Sc$ ultimately grows as $\rho^2\gamma$ and not
$(\rho\gamma)^2$ as thought previously \cite{GW}.

A pair collision theory is strictly valid only when the mean free path
is large compared to $\rc$, when effects such as finite collision
volumes and correlations are unimportant. The excellent comparison with
simulation implies though that the transport coefficients are rather
insensitive to these effects so long as $\lambda\agt1$. There must be a
point though where these effects become important, and we now turn to a
discussion of this, the large $\rho$ limit.

Firstly, note that from a kinetic theory point of view the large $\rho$
limit is rather unusual. For hard spheres for instance there is a
natural maximum density, but for the ideal dissipative fluid, $\rho$ can
be increased indefinitely. In this limit each sphere is in continual
interaction with many others simultaneously and a picture in terms of
pair collisions is surely invalid. Moreover, one can derive some of the
MBE results by assuming a mean field interaction of this type \cite{GW}.
Therefore at very high densities, one would expect the MBE results (for
$\tau$ and $D$ at least) to be recovered. A slightly more sophisticated
approach would couple the motion of an individual particle to the
hydrodynamic modes of the fluid, and recover for instance the celebrated
long time tail in the VACF. An elegant mode-coupling theory along these
lines has recently been developed by Espa\~nol and coworkers
\cite{Esp2}.

Fig.~2(a) shows how $\etaK$ starts to drift down from the Boltzmann pair
collision theory toward the MBE result for densities $\rho\agt10$. It is
at first sight remarkable that a pair collision theory works at such
high densities. We can partially understand this observation though when
we consider the fluctuations in $\tau^{-1}$ in Eq.~\eqref{taueq},
interpreting the integral as a drag-per-particle in a random stationary
background of other particles. If we suppose that the number of
particles in a volume element $d^3\r$ is a Poisson distribution with
mean $\rho\,d^3\r$ and variance equal to the mean, then 
$\tau^{-1}$ is a weighted sum of independent random variables whose
mean is given by Eq.~\eqref{taueq} and variance by
\begin{equation}
\var\tau^{-1}=\frac{1}{3}\int\!\rho\,d^3\r\,[\gamma\wD(r)]^2
=\frac{4\pi\gamma^2\rho}{315}.
\end{equation}
Thus the ratio of the standard deviation to the mean for $\tau^{-1}$ is
$(\rho/\rhoC)^{-1/2}$ where $\rhoC=45/7\pi$. For $\rho=10$, for
instance, this ratio is about 45\%, and to get the ratio $\alt10$\%
requires $\rho\agt200$. Thus unless $\rho\gg1$ a particle sees large
fluctuations in its interaction with other particles ($O(1)$ in
essence), and it is perhaps not surprising that a pair collision theory
is applicable even though the apparent density becomes rather large.
Another way of stating this conclusion is to say that $\rc^3$ 
severely over-estimates the effective interaction volume.

The above discussions have concerned the ideal dissipative fluid. Let us
finish off by briefly discussing the properties of the non-ideal fluid.
The main effect of the soft repulsion forces is to provide another
collision mechanism which is most significant at low $\gamma$; in fact
at $\gamma=0$ one recovers a regular fluid of soft spheres. The effect
is to remove the $1/\gamma$ divergence in the kinetic coefficients as
$\gamma\to0$. Combined with the plateau that is expected for
$\gamma\agt\gammaC$, this makes the transport coefficients very
insensitive to parameter variations. The collision integrals can also be
evaluated for non-ideal interactions, and again we find excellent
agreement with simulations. This is so even though the VACF in some
instances develops a pronounced non-trivial structure.

In summary therefore, we have argued that, for the typical parameters
used in applications, the kinetic transport coefficients in DPD are best
captured by a Boltzmann pair collision theory, even though the apparent
density is rather high. The evidence is the excellent agreement with
simulation results, and the fact that the number of collisions a
particle experiences in a mean free path appears to be the principal
parameter that determines the kinetic properties. The kinetic theory of
Marsh et al \cite{MBE} is recovered as a weak scattering or Vlasov
theory, valid in the low $\gamma$ limit. Previously reported
discrepancies with this theory \cite{MBE,PHF} are thereby resolved. We
further argue that a qualitative change in the scaling of kinetic
transport coefficients has to occur at large $\gamma$ since at least one
collision per mean free path is required to decorrelate a particle's
initial velocity. This conclusion, too, is supported by our simulation
evidence. Finally, the present study is likely to have implications for
the analysis of transport behaviour in the various generalisations of
DPD \cite{DPDgen}. More details of our results and our parallel
investigations of the non-ideal fluid will be the subject of a longer
paper.

\stars

We acknowledge useful discussions and correspondence with {\sc P.
Espa\~nol}, {\sc M. H. J. Hagen}, {\sc C. Lowe} and {\sc I.
Pagonabarraga}. AJM would like to thank the British Council and the
Nederlandse Organisatie voor Wetenschappelijk Onderzoek (NWO) for
providing a travel grant which aided this work enormously.

\newpage

%
%

\begin{figure}
\centering
\includegraphics{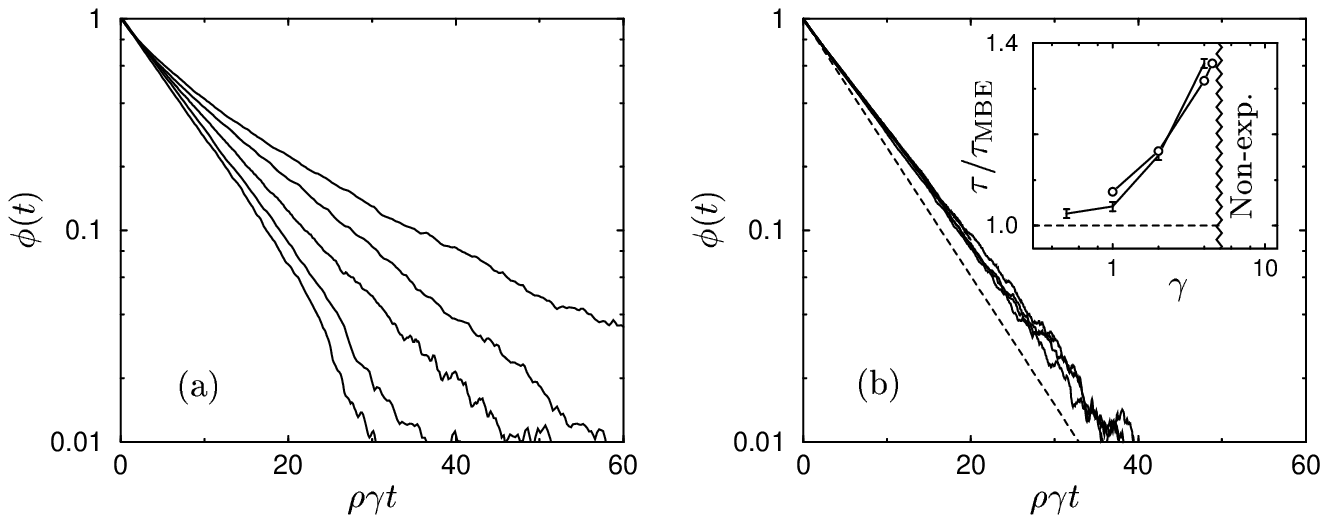}
\caption[?]{Velocity autocorrelation function for ideal dissipative
fluid: (a) at $\rho=3$ for $\gamma=1$ (steepest curve), 2, 4, 8, 16, and
(b) at $\gamma=2$ for $\rho=1$, 2, 3, 4, 5, 6, showing data collapse
(note that a basic scaling prefactor $\rho\gamma$ has been extracted from
the time dependence). The dashed line in (b) shows the universal decay
predicted by Marsh et al \cite{MBE} (MBE). Where single exponential
decay occurs over two decades or more, the inset to (b) shows the decay
rate normalised by the MBE result, $\tau_{\mathrm{MBE}} =
45/2\pi\rho\gamma$, (points with error bars), as a function of $\gamma$.
The open circles in this plot are predictions from the Boltzmann pair 
collision theory discussed in the text.}
\end{figure}

\begin{figure}
\centering
\includegraphics{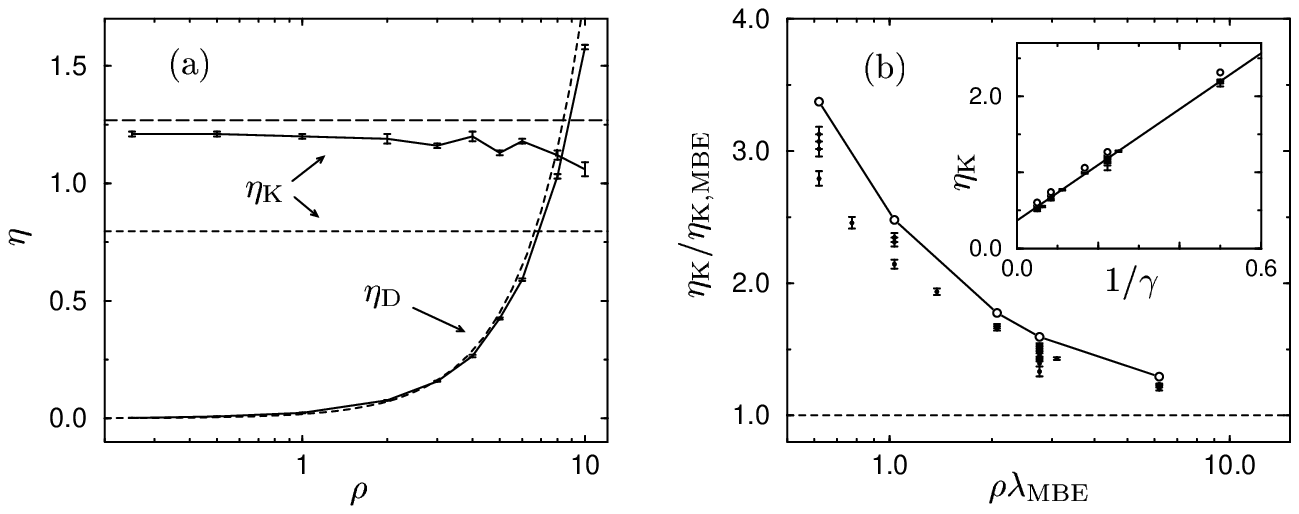}
\caption[?]{Viscosity of ideal dissipative fluid: (a) the kinetic and
dissipative contributions, $\etaK$ and $\etaD$, at $\gamma=4.5$ as a
function of density, and (b) data collapse for all $\etaK$ data
($\rho=0.25$--10, $\gamma=2$--20), normalised with the MBE result,
$\eta_{\mathrm{K,MBE}}=45/4\pi\gamma$, and plotted against
$\rho\lambdaMBE=45\sqrt{3}/2\pi\gamma$. The dashed lines in (a) and (b) are
the predictions of MBE theory. The long dashed line in (a) and the open
circles in (b) are predictions from the Boltzmann pair collision theory
in the text. The inset in (b) shows all the $\etaK$ data plotted against
$1/\gamma$ with a linear regression line. Note the non-zero intercept in
this plot, which indicates a plateau in $\etaK$ for large $\gamma$.}
\end{figure}


\vskip-12pt
\begin{thebibliography}{99}

\bibitem{DPD}{\Name{Boek E. S., Coveney P. V., Lekkerkerker H. N. W.
\And van der Schoot P.} \Review{Phys. Rev. E} \Vol{55} \Year{1997}
\Page{3124}; \Name{Novik K. E. \And Coveney P. V.} \Review{Int. J. Mod.
Phys. C} \Vol{1997} \Year{1997} \Page{909}; \Name{Kong Y., Manke C. W.,
Madden W. G. \And Schlijper A. G.} \Review{J. Chem. Phys.} \Vol{107}
\Year{1997} \Page{592}; \Name{Groot R. D. \And Madden T. J.} \Review{J.
Chem. Phys.} \Vol{108} \Year{1998} \Page{8713}. For a recent review of
the area see \Name{Warren P. B.} \Review{Current Opinion Coll. Int.
Sci.} \Vol{3} \Year{1998} \Page{620}.}

\bibitem{HK}{\Name{Hoogerbrugge P. J. \And Koelman J. M. V. A.}
\Review{Europhys. Lett.} \Vol{19} \Year{1992} \Page{155}.}

\bibitem{EW}{\Name{Espa\~nol P. \And Warren P. B.}
\Review{Europhys. Lett.} \Vol{30} \Year{1995} \Page{191}.}

\bibitem{GW}{\Name{Groot R. D. \And Warren P. B.} \Review{J. Chem. Phys.}
\Vol{107} \Year{1998} \Page{4423}.}

\bibitem{Esp}{\Name{Espa\~nol P.} \Review{Phys. Rev. E} \Vol{52}
\Year{1995} \Page{1734}.}

\bibitem{MBE}{\Name{Marsh C. A., Backx G. \And Ernst M. H.}
\Review{Europhys. Lett.} \Vol{38} \Year{1997} \Page{411}; \Review{Phys.
Rev. E} \Vol{56} \Year{1997} \Page{1676}.}

\bibitem{PHF}{\Name{Pagonabarraga I., Hagen M. H. J \And Frenkel D.}
\Review{Europhys. Lett.} \Vol{42} \Year{1998} \Page{377}.  
The two dimensional $\etaK$ data of these authors also shows a
scaling collapse similar to Fig.~2(b) in the present work, if plotted
against $\rho\lambdaMBE$ (\Name{I. Pagonabarraga} private communication).}

\bibitem{sims}{For algorithm details refer to \cite{GW}. We have been
careful to keep systematic artifacts such as those due to a finite box
size, finite shear rate or a finite time step size to within the error
bars. All simulations were done in three dimensions. Shear was applied
using Lees-Edwards sliding boundary conditions.}

\bibitem{LTT}{\Name{Alder B. J. \And Wainwright T. E.} \Review{Phys.
Rev.} \Vol{A1} \Year{1970} \Page{18}. Our simulations were intended to
probe the early time behaviour of the VACF, and we have not carried them
through to sufficient precision to see the long time tail unambiguously.}

\bibitem{Res}{\Name{Chapman S. \And Cowling T.} \Review{The mathematical
theory of non-uniform gases} (CUP, Cambridge) \Year{1939};
\Name{R\'esibois P. \And de Leener M.} \Review{Classical kinetic theory
of fluids} (Wiley, New York) \Year{1977}.}

\bibitem{Evans}{\Name{Evans G. T.} \Review{J. Chem. Phys.} \Vol{110}
\Year{1999} \Page{1338}.}

\bibitem{Esp2}{\Name{Espa\~nol P.} private communication.}

\bibitem{DPDgen}{\Name{Avalos J. B. \And Mackie A. D.} \Review{Europhys.
Lett.} \Vol{40} \Year{1997} \Page{141}; \Name{Espa\~nol P.}
\Review{Europhys. Lett.} \Vol{40} \Year{1997} \Page{631};
\Name{Espa\~nol P.} \Review{Phys. Rev. E} \Vol{57} \Year{1998}
\Page{2930}.}

\end{thebibliography}
\end{document}